\documentclass[aps,prl,twocolumn,superscriptaddress,preprintnumbers]{revtex4-1}
\usepackage{graphicx}
\usepackage{amsmath}
\usepackage{amssymb}
\usepackage{bm}
\usepackage{epsfig}
\begin{document}

\title{\boldmath 
\hspace{-4ex}
Inclusive Hadroproduction of $P$-Wave Heavy Quarkonia in Potential
Nonrelativistic QCD}
\preprint{TUM-EFT 138/20}

\author{Nora~Brambilla}
\affiliation{Physik-Department, Technische Universit\"at M\"unchen, James-Franck-Str. 1, 85748 Garching, Germany}
\affiliation{Institute for Advanced Study, Technische Universit\"at M\"unchen, Lichtenbergstrasse 2~a, 85748 Garching, Germany}
\author{Hee~Sok~Chung}
\affiliation{Physik-Department, Technische Universit\"at M\"unchen, James-Franck-Str. 1, 85748 Garching, Germany}
\author{Antonio~Vairo}
\affiliation{Physik-Department, Technische Universit\"at M\"unchen, James-Franck-Str. 1, 85748 Garching, Germany}

\date{\today}

\begin{abstract}
We compute the color-singlet and color-octet nonrelativistic QCD (NRQCD) 
long-distance matrix elements for inclusive production of $P$-wave quarkonia in
the framework of potential NRQCD. 
In this way, the color-octet NRQCD long-distance matrix element can be determined without relying on measured cross section data, which has not been possible so far. 
We obtain inclusive cross sections of $\chi_{cJ}$ and $\chi_{bJ}$ at the LHC, which are in good agreement with data. 
In principle, the formalism developed in this Letter can be applied to all inclusive production processes of heavy quarkonia. 
\end{abstract}

\maketitle

The mechanism underlying heavy quarkonium production is a key to understanding the dynamics of strongly coupled
systems~\cite{Brambilla:2004wf,Brambilla:2010cs,Bodwin:2013nua,Brambilla:2014jmp,Lansberg:2019adr}. 
Quarkonium production is extensively studied in experiments at particle colliders like LHC, SuperKEKB, BEPC II, and RHIC,
and will continue to be an important subject in future colliders such as the planned electron-ion collider.
Quarkonium production has a large impact on studies of the QCD phase diagrams and early Universe,
as the production in proton-proton collisions is the bottom line to which quarkonium suppression in heavy ion collisions is compared~\cite{Matsui:1986dk}. 
Moreover, from the theoretical point of view, quarkonium production processes have exquisite theoretical issues
pinning down factorization in strongly coupled theories, definition and calculation of nonperturbative matrix elements, 
and resummation of logarithms of large ratios of scales~\cite{Bodwin:1994jh,Nayak:2005rw,Nayak:2005rt,Nayak:2006fm,Bodwin:2019bpf}. 

The typical hierarchy of energy scales that characterizes heavy quarkonium is $m \gg mv \gg mv^2$,
where $m$ is the heavy quark mass and $v \ll 1$ is the relative velocity of the quark in the bound state.
This hierarchy of energy scales may be exploited to construct a hierarchy of effective field theories. 
Nonrelativistic QCD (NRQCD)~\cite{Caswell:1985ui,Bodwin:1994jh} follows from QCD by integrating out
modes associated with the energy scale $m$ from Green's functions describing a heavy quark and a heavy antiquark near threshold.
The matching to NRQCD can be done perturbatively, since $m$ is larger than the typical hadronic scale $\Lambda_{\rm QCD}$. 
Potential NRQCD (pNRQCD)~\cite{Pineda:1997bj,Brambilla:1999xf,Brambilla:2004jw} follows from NRQCD by integrating out gluons of energy or momentum of order $mv$.
The matching to pNRQCD may need to rely on nonperturbative methods if the momentum scale $mv$ is comparable to~$\Lambda_{\rm QCD}$.

While NRQCD had great success in heavy quarkonium phenomenology, a satisfactory description of inclusive production processes from first principles is still beyond reach. 
Much of the difficulty stems from our limited knowledge of the NRQCD long-distance matrix elements (LDMEs),
which describe the nonperturbative evolution of the heavy quark $Q$ and antiquark $\bar Q$ into a quarkonium. 
First-principles determinations have not been possible, even approximately, for a class of important LDMEs that are associated with the $Q \bar Q$ in a 
color-octet state. 
On the other hand, phenomenological determinations of the unknown LDMEs based on different choices of observables have led to inconsistent sets of LDMEs,
which have resulted in contradicting predictions, in particular, leaving open the long-standing problem of the polarization of quarkonium produced in hadron colliders~\cite{Chung:2018lyq}. 
It would be of enormous impact to be able to compute the unknown LDMEs from first principles. 

Potential NRQCD has been successfully applied to annihilation and exclusive electromagnetic production processes of heavy quarkonia~\cite{Brambilla:2001xy,Brambilla:2002nu,Brambilla:2020xod}. 
It has been anticipated that pNRQCD could also be used to describe inclusive production processes.
In this Letter, we apply for the first time pNRQCD to these kinds of processes by computing 
the NRQCD LDMEs that appear in the inclusive production cross section of $P$-wave quarkonia. 
Specifically, we consider production cross sections of $\chi_{QJ}$  ($Q = c$ or $b$, $J=0$, 1, and 2) at leading order in $v$. 

The cross section is given in the NRQCD factorization formalism at leading order in $v$ by~\cite{Bodwin:1994jh}
\begin{align}
\label{eq:NRQCDfac}
\sigma_{\chi_{QJ}+X} &=  (2 J+1) \sigma_{Q \bar Q({}^3P_J^{[1]})} \langle {\cal O}^{\chi_{Q0}} ({}^3P_0^{[1]}) \rangle 
\nonumber \\
& \quad + (2 J+1) \sigma_{Q \bar Q({}^3S_1^{[8]})} \langle {\cal O}^{\chi_{Q0}} ({}^3S_1^{[8]}) \rangle. 
\end{align}
Here, we use spectroscopic notation for the angular momentum state of the $Q
\bar Q$, while the superscripts 1 and 8 denote the color state of the $Q \bar Q$: color singlet (CS) and color octet (CO), respectively.
The quantities $\sigma_{Q \bar Q({}^3P_J^{[1]})}$ and $\sigma_{Q \bar Q({}^3S_1^{[8]})}$ are the perturbatively calculable short-distance coefficients (SDCs).
We have used the heavy quark spin symmetry to reduce the $\chi_{QJ}$ LDMEs into LDMEs of $\chi_{Q0}$, which are defined by
\begin{subequations}
\label{eq:matrix_elements}
\begin{align}
\langle {\cal O}^{\chi_{Q0}} ({}^3P_0^{[1]}) \rangle &= \frac{1}{3} \langle \Omega | \chi^\dag (- \tfrac{i}{2} \overleftrightarrow{\bm{D}} \cdot \bm{\sigma}) \psi {\cal P}_{\chi_{Q0}(\bm{P}=\bm{0})} 
\nonumber \\ & \quad \times \psi^\dag (- \tfrac{i}{2} \overleftrightarrow{\bm{D}} \cdot \bm{\sigma}) \chi | \Omega \rangle, 
\\
\langle {\cal O}^{\chi_{Q0}} ({}^3S_1^{[8]}) \rangle &= \langle \Omega | \chi^\dag \sigma^i T^a \psi \Phi_\ell^{\dag ab} {\cal P}_{\chi_{Q0}(\bm{P}=\bm{0})} 
\nonumber \\ & \quad \times \Phi_\ell^{bc} \psi^\dag \sigma^i T^c \chi | \Omega \rangle, 
\end{align}
\end{subequations}
where $| \Omega \rangle$ is the QCD vacuum, $T^a$ are SU(3) generators,
$\sigma^i$ are Pauli matrices, 
and $\psi$ and $\chi$ are Pauli spinors that annihilate and create a heavy quark and antiquark, respectively. 
The covariant derivative $\overleftrightarrow{\bm{D}}$ is defined by $\chi^\dag \overleftrightarrow{\bm{D}} \psi = \chi^\dag \bm{D} \psi - (\bm{D} \chi)^\dag \psi$, 
with $\bm{D} = \bm{\nabla} -i g \bm{A}$, and $\bm{A}$ is the gluon field. 
The operator ${\cal P}_{{\cal Q} (\bm{P})}$ projects onto a state consisting of a quarkonium $\cal Q$ with momentum $\bm{P}$.
The path-ordered Wilson line along the spacetime direction $\ell$, defined by $\Phi_\ell = {\cal P} \exp [ -i g \int_0^\infty d\lambda\, \ell \cdot A^{\rm adj}(\ell \lambda) ]$,
where $A^{\rm adj}$ is the gluon field in the adjoint representation, ensures the gauge invariance of the CO LDME~\cite{Nayak:2005rw}. 
The direction $\ell$ is arbitrary. 

We aim at expressing the LDMEs \eqref{eq:matrix_elements} in pNRQCD.
We work in the strong coupling regime, where $\Lambda_{\rm QCD} \gg mv^2$.
This condition is fulfilled by non-Coulombic quarkonia. 
Possible non-Coulombic quarkonia are excited states such as the 
$\chi_{QJ}$ that we consider in this Letter.
In order to compute the LDMEs in strongly coupled pNRQCD,
we expand the NRQCD Hamiltonian in inverse powers of the heavy quark mass, 
\begin{equation}
\label{QMPT1}
H_{\rm NRQCD} = H_{\rm NRQCD}^{(0)} + \frac{H_{\rm NRQCD}^{(1)}}{m} + \cdots ,
\end{equation}
and compute at each order in $1/m$ in quantum-mechanical perturbation theory (QMPT) its eigenstates~\cite{Brambilla:2000gk,Pineda:2000sz},
\begin{equation}
\label{QMPT2}
|\underline{n}; \bm{x}_1, \bm{x}_2 \rangle = |\underline{n}; \bm{x}_1, \bm{x}_2 \rangle^{(0)} + \frac{|\underline{n}; \bm{x}_1, \bm{x}_2 \rangle^{(1)}}{m} + \cdots. 
\end{equation}
Here $\bm{x}_1$ and $\bm{x}_2$ are the positions of the quark and antiquark, respectively.
Only for the eigenstates in the static limit, $ |\underline{n}; \bm{x}_1, \bm{x}_2 \rangle^{(0)}$, the quark and antiquark positions are conserved. 
The corresponding eigenvalues $E_n^{(0)}(\bm{x}_1, \bm{x}_2)$ are the energies of a static quark-antiquark pair
located in $\bm{x}_1$ and $\bm{x}_2$ in the presence of light degrees of freedom in the ground state ($n=0$) or in some excited state ($n>0$).
The set of quantum numbers denoted with $n$ labels the excitations of the light degrees of freedom. 
We also define the states $|n; \bm{x}_1, \bm{x}_2 \rangle$, which encode the light degrees of freedom content of the states $|\underline n; \bm{x}_1, \bm{x}_2 \rangle$, 
through the relation $| \underline n; \bm{x}_1, \bm{x}_2 \rangle = \psi^\dag(\bm{x}_1) \chi(\bm{x}_2) | n; \bm{x}_1, \bm{x}_2 \rangle$.

The computation of the LDMEs in Eqs.~\eqref{eq:matrix_elements} requires the knowledge of the matrix elements of ${\cal P}_{{\cal Q}(\bm{P}=\bm{0})}$ 
on the states $|\underline n; \bm{x}_1, \bm{x}_2 \rangle$, which follows from general properties of the operator.
We first note that ${\cal P}_{{\cal Q}(\bm{P})}$ commutes with the NRQCD Hamiltonian.
This is because the decay modes that involve momentum transfers at the scale $m$ have been integrated out,
and hadronic and electromagnetic transitions between quarkonium states can be ignored as they occur at a much larger timescale than the hadronization of $Q$ and $\bar Q$ into quarkonium.
Hence, the quarkonium state $\cal Q$ is conserved and so is ${\cal P}_{{\cal Q}(\bm{P})}$.
We write an eigenstate of $H_{\rm NRQCD}$ and ${\cal P}_{{\cal Q}(\bm{P})}$ as
\begin{equation}
\label{eq:quarkonium_state}
|{\cal Q}(n,\bm{P})\rangle = \int d^3x_1 d^3x_2 \, \phi_{{\cal Q}(n,\bm{P})}(\bm{x}_1,\bm{x}_2) \, |\underline{n}; \bm{x}_1, \bm{x}_2 \rangle, 
\end{equation}
where $\phi_{{\cal Q}(n,\bm{P})}(\bm{x}_1,\bm{x}_2)$ is a suitable function of $\bm{x}_1$, $\bm{x}_2$, and $\bm{P}$.
We take $|{\cal Q}(n,\bm{P})\rangle$ to be nonrelativistically normalized. 
For $n=0$, the state is just a quarkonium state $|{\cal Q}(\bm{P}) \rangle$~\cite{Brambilla:2002nu}.
For $n \neq 0$, the state includes excitations of the light degrees of freedom. 

For the state $|\underline{n};\bm{x}_1,\bm{x}_2\rangle$ to have a nonvanishing overlap with ${\cal P}_{{\cal Q}(\bm{P})}$,
its static component $|\underline{n};\bm{x}_1,\bm{x}_2\rangle^{(0)}$ must be of the form
${\rm Tr}\{\psi^\dag(\bm{x}_1) \chi(\bm{x}_2)\}\,{\rm Tr}\{| n; \bm{x}_1, \bm{x}_2 \rangle^{(0)}\}/N_c$
in the limit  $\bm{x}_1 -\bm{x}_2 \to \bm{0}$, ${\rm Tr}$ being the trace over color and $N_c$ the number of colors~\footnote{
In the special case $n=0$, it holds that \unexpanded{${\rm Tr}\{| 0; \bm{x}_1, \bm{x}_2 \rangle^{(0)}\} = \sqrt{N_c} |\Omega\rangle$}
for $\bm{x}_1 -\bm{x}_2 \to \bm{0}$~\cite{Brambilla:2002nu,Brambilla:2020xod}.}.  
This condition follows from requiring that, in the absence of gluonic excitations carried by $1/m$ corrections,
the quark-antiquark fields at the same point are in a color-singlet configuration,
which, in turn, is a necessary requirement for the state  $|\underline{n};\bm{x}_1,\bm{x}_2\rangle$ to overlap with a state containing a quarkonium.
We denote with $\mathbb S$ the subset of eigenstates of the NRQCD Hamiltonian $|\underline{n};\bm{x}_1,\bm{x}_2\rangle$ that fulfill the condition.
Finally, the operator ${\cal P}_{{\cal Q} (\bm{P})}$ can be written as
\begin{align}
\label{eq:projection_NRQCD}
&{\cal P}_{{\cal Q}(\bm{P})} = \sum_{n\in\mathbb{S}} |{\cal Q}(n,\bm{P})\rangle  \langle{\cal Q}(n,\bm{P})|. 
\end{align}

In strongly coupled pNRQCD~\cite{Brambilla:1999xf,Brambilla:2000gk,Pineda:2000sz}, color-singlet heavy quark-antiquark pairs
in the presence of light degrees of freedom in a state $n$ are described by the field $S_n$ and the Hamiltonian
\begin{equation}
\label{eq:pNRQCD_hamiltonian}
H_{\rm pNRQCD} = \int d^3x_1 d^3x_2 \, S_n^\dag h_n (\bm{x}_1, \bm{x}_2;\bm{\nabla}_1, \bm{\nabla}_2 ) S_n, 
\end{equation}
where $h_n\delta^{(3)} (\bm{x}_1-\bm{x}_1')\delta^{(3)} (\bm{x}_2-\bm{x}_2')$ is determined
by matching order by order in $1/m$ to $\langle \underline{n}; \bm{x}_1, \bm{x}_2 |H_{\rm NRQCD}| \underline{n}; \bm{x}'_1, \bm{x}'_2 \rangle$.
In particular, at leading order in $v$, $h_n$ is the sum of a kinetic energy operator $\bm{p}^2/m$,
where $\bm{p} =  -i(\bm{\nabla}_1-\bm{\nabla}_2)/2$ is the relative momentum of the heavy quark-antiquark pair,
and a static potential that matches the eigenvalue $E^{(0)}_n$ of $H^{(0)}_{\rm NRQCD}$.

Since the eigenvalues of $h_n$ have to be equal to $\langle {\cal Q}(n,\bm{P})| H_{\rm NRQCD}|{\cal Q}(n,\bm{P})\rangle$,
as pNRQCD and NRQCD describe the same physical spectrum, we can identify the functions $\phi_{{\cal Q}(n,\bm{P})}(\bm{x}_1,\bm{x}_2)$ with the eigenfunctions of $h_n$.
At leading order in $v$, it holds
\begin{align}
\label{eq:pNRQCD_wavefunction}
\phi_{{\cal Q}(n,\bm{P})}(\bm{x}_1,\bm{x}_2) \approx e^{-i\bm{P}\cdot(\bm{x}_1+\bm{x}_2)/2}\phi^{(0)}_{{\cal Q}(n)} (\bm{x}_1,\bm{x}_2),
\end{align}
where ${\rm exp}[-i\bm{P}\cdot(\bm{x}_1+\bm{x}_2)/2]$ is a plane wave describing the center of mass motion and
$\phi^{(0)}_{{\cal Q}(n)}(\bm{x}_1,\bm{x}_2)$ is an eigenfunction of the leading-order pNRQCD Hamiltonian.

We are now in the position to express the production LDMEs in terms of gluon field correlators and the wave functions at the origin.
The procedure is similar to the one developed to compute the annihilation rates of heavy quarkonia in pNRQCD in Refs.~\cite{Brambilla:2001xy,Brambilla:2002nu,Brambilla:2020xod}
and consists of the following steps:
{\it (i)} Replace in the LDMEs the projector ${\cal P}_{{\cal
Q}(\bm{P}=\bm{0})}$ with the expressions \eqref{eq:projection_NRQCD}
and \eqref{eq:quarkonium_state}.
{\it (ii)} Using QMPT and, in particular, Eqs. \eqref{QMPT1} and \eqref{QMPT2},
express the LDMEs in terms of $|\underline{n}; \bm{x}_1, \bm{x}_2 \rangle^{(0)}$ and $E_n^{(0)}(\bm{x}_1, \bm{x}_2)$.
{\it (iii)} Make explicit the heavy quark and antiquark field content of the states $|\underline{n}; \bm{x}_1, \bm{x}_2 \rangle^{(0)}$ and eliminate the fields by using Wick's theorem;
one makes use at this point of the fact that the states in ${\cal P}_{{\cal Q}(\bm{P}=\bm{0})}$ belong to the set $\mathbb S$, which constrains their color structure.
{\it (iv)} Rewrite the sum of the matrix elements of the gluon fields on the states $|\Omega \rangle$ and $|n; \bm{x}_1, \bm{x}_2 \rangle^{(0)}$ (evaluated at $\bm{x}_1-\bm{x}_2=\bm{0}$) 
in terms of gluon field correlators.
{\it (v)} Identify  $\phi_{{\cal Q}(n,\bm{P})}(\bm{x}_1,\bm{x}_2)$ or derivatives of them (evaluated at $\bm{x}_1-\bm{x}_2=\bm{0}$) with the wave functions of the pNRQCD Hamiltonian.
The leading-order wave functions can be computed by solving the corresponding Schr\"odinger equations, 
once the static potentials have been determined from the static energies $E^{(0)}_n$ typically obtained by lattice QCD methods. 

The wave functions $\phi^{(0)}_{{\cal Q}(n)} (\bm{x}_1,\bm{x}_2)$ depend on $n$ and may differ from the usual quarkonium wave functions
$\phi^{(0)}_{{\cal Q}} (\bm{x}_1,\bm{x}_2)$ that correspond to the $n=0$ case.
For $n=0$, the static potential in $h_0$ can be extracted from the vacuum expectation value (VEV) of a static Wilson loop~\cite{Brambilla:1999xf,Brambilla:2000gk}. 
Similarly, for $n \neq 0$ and $n \in \mathbb S$,
the static potential in $h_n$ can be extracted from the VEV of a static Wilson loop in the presence of some additional, disconnected gluon fields.
To our knowledge, there are no lattice data available for the $n \neq 0$ static potentials.
However, we expect that the disconnected gluon fields mostly provide a constant shift to the potentials, for instance in the form of a glueball mass,
but do not significantly affect their slopes.
This is also supported by large $N_c$ considerations.
In the large $N_c$ limit, the VEV of a Wilson loop with additional disconnected gluon fields factorizes into the VEV of the Wilson loop times the VEV of the additional gluon fields
up to corrections of order $1/N_c^2$~\cite{Makeenko:1979pb, Witten:1979pi}. 
If the slopes of the static potentials are the same for all $n$ in the large $N_c$ limit,
then in that limit the wave functions $\phi^{(0)}_{{\cal Q}(n)} (\bm{x}_1,\bm{x}_2)$ are independent of $n$.
Hence, we will approximate the wave functions $\phi^{(0)}_{{\cal Q}(n)} (\bm{x}_1,\bm{x}_2)$ with the quarkonium wave function $\phi^{(0)}_{{\cal Q}}(\bm{x}_1,\bm{x}_2)$
making an error of at most $O(1/N_c^2)$. 

Following the outlined procedure,
we can compute the production LDMEs $\langle {\cal O}^{\chi_{Q0}} ({}^3P_0^{[1]}) \rangle$ and $\langle {\cal O}^{\chi_{Q0}} ({}^3S_1^{[8]}) \rangle$ in strongly coupled pNRQCD. 
Furthermore, in the case of the CO LDME, we approximate $\phi^{(0)}_{{\chi_{Q0}}(n)} (\bm{x}_1,\bm{x}_2) \approx \phi^{(0)}_{{\chi_{Q0}}} (\bm{x}_1,\bm{x}_2)$ as discussed above.

For the CS LDME $\langle {\cal O}^{\chi_{Q0}} ({}^3P_0^{[1]}) \rangle$, we obtain at leading order in QMPT
\begin{equation}
\label{eq:singlet_result}
\langle {\cal O}^{\chi_{Q0}} ({}^3P_0^{[1]}) \rangle  = \frac{3 N_c}{2 \pi} | R^{(0)'}_{\chi_{Q0}}(0)|^2, 
\end{equation}
where $R_{\chi_{Q0}}^{(0)}(r)$ is the radial wave function of $\chi_{Q0}$ at leading order in the velocity expansion
[$R^{(0)'}_{\chi_{Q0}}(r)$ stands for its derivative]. 
This reproduces the result obtained in the vacuum-saturation approximation in Ref.~\cite{Bodwin:1994jh}. 

The CO LDME $\langle {\cal O}^{\chi_{Q0}} ({}^3S_1^{[8]}) \rangle$ vanishes at leading order in QMPT.
Nonvanishing contributions come from next-to-leading order in QMPT,
\begin{align}   
\label{eq:octet_result}
&\langle {\cal O}^{\chi_{Q0}} ({}^3S_1^{[8]}) \rangle = \frac{3 N_c}{2 \pi} | R^{(0)'}_{\chi_{Q0}}(0)|^2 \, \frac{{\cal E}}{9 N_c m^2},
\end{align}
where
\begin{align}
&{\cal E} = \frac{3}{N_c} \int_0^\infty dt\, t \int_0^\infty dt'\, t'
\nonumber \\
& \times \langle \Omega | \Phi_\ell^{\dag ab} \Phi_0^{\dag da} (0,t) g E^{d,i} (t) g E^{e,i} (t') \Phi^{ec}_0 (0,t') \Phi_\ell^{bc} | \Omega \rangle, 
\end{align}
$E^{a,i}(t)$ being a chromoelectric field component computed at the time $t$ and at the space location $\bm{0}$, 
and $\Phi_0 (t,t') = {\cal P} \exp [ -i g \int_t^{t'} d\tau\, A_0^{\rm
adj}(\tau,\bm{0}) ]$ being a Schwinger line.
Note that ${\cal E}$ is a purely gluonic quantity that does not depend on the heavy quark flavor. 

The expression for the CO LDME given in Eq.~\eqref{eq:octet_result} is very similar to the pNRQCD expression for the CO LDME appearing at leading order in $v$ 
in the decay of $\chi_{QJ}$ into light hadrons~\cite{Brambilla:2001xy,Brambilla:2020xod}.
The only difference is that there the correlator ${\cal E}$ is replaced by the correlator 
\begin{equation}
\label{eq:correlator4}
{\cal E}_3 = \frac{1}{2N_c} \int_0^\infty dt \, t^3 \,\langle \Omega | g E^{a,i} (t) \Phi_0^{ab}(t,0)g E^{b,i} (0) | \Omega \rangle.
\end{equation}
The two correlators would be the same if we could neglect the contributions from the strings.
At one loop, they have the same logarithmic dependence on the renormalization scale $\Lambda$ and satisfy the same evolution equation.
From this~\cite{Brambilla:2001xy}, it follows that the one-loop evolution equation for $\langle {\cal O}^{\chi_{Q0}} ({}^3S_1^{[8]}) \rangle$ is 
\begin{equation}
\label{eq:RG}
\frac{d}{d \log \Lambda} \langle {\cal O}^{\chi_{Q0}} ({}^3S_1^{[8]}) \rangle = \frac{4 C_F \alpha_s}{3 N_c \pi m^2} \langle {\cal O}^{\chi_{Q0}} ({}^3P_0^{[1]}) \rangle,
\end{equation}
where $C_F = (N_c^2-1)/(2 N_c)$. 
Equation~\eqref{eq:RG} agrees with the evolution equation derived from a perturbative calculation in NRQCD~\cite{Bodwin:1994jh}.
The agreement is a one-loop consistency check of Eq.~\eqref{eq:octet_result}.
At two loops, however, the identification of ${\cal E}$ with ${\cal E}_3$ may not hold~\cite{Nayak:2005rw,Nayak:2005rt,Nayak:2006fm,Bodwin:2019bpf}.

Equation~\eqref{eq:octet_result} is our result for the CO LDME. 
The result allows a first-principles determination of the CO LDME, once ${\cal E}$ is known.
The correlator ${\cal E}$ may be computed in lattice QCD or it can be obtained from processes involving heavy quarkonia.

We now compute the inclusive production cross sections of $\chi_{cJ}$ and $\chi_{bJ}(nP)$ from proton-proton collisions at the LHC
based on our results for the LDMEs in  Eqs.~\eqref{eq:singlet_result} and \eqref{eq:octet_result}.
We use the value $|R^{(0)'}_{\chi_{c0}}(0)|^2 = 0.057$~GeV$^5$, which we obtain by comparing the measured two-photon decay rates of $\chi_{c0}$ and $\chi_{c2}$ in Ref.~\cite{Ablikim:2012xi}
with the pNRQCD expressions at leading order in $v$ and at next-to-leading order (NLO) in $\alpha_s$~\cite{Brambilla:2020xod}. 
Because two-photon decay rates of $\chi_{bJ}$ have not been measured yet, 
we take for $|R^{(0)'}_{\chi_{b0}(nP)}(0)|^2$ the averages of the values listed in Ref.~\cite{Brambilla:2020xod}, 
which are obtained from the potential-model calculations in Refs.~\cite{Buchmuller:1980su,Eichten:1995ch,Chung:2010vz,Eichten:2019hbb}. 
We take $|R^{(0)'}_{\chi_{b0}(1P)}(0)|^2 = 1.47~$~GeV$^5$, $|R^{(0)'}_{\chi_{b0}(2P)}(0)|^2 = 1.74~$~GeV$^5$, and $|R^{(0)'}_{\chi_{b0}(3P)}(0)|^2 = 1.92~$~GeV$^5$.
We take $m$ to be 1.5~GeV for charm and 4.75~GeV for bottom.
The LDMEs are completely fixed by $|R^{(0)'}_{\chi_{Q0}}(0)|$, ${\cal E}$, and $m$. 
We include the uncertainties coming from ${\cal E}$ and from unaccounted relativistic corrections of relative order $v^2$,
which we take to be 30\% (10\%) of the central values of the $\chi_{cJ}$ ($\chi_{bJ}$) cross sections. 
We neglect the uncertainty from the corrections of order $1/N_c^2$, which is small compared to other uncertainties. 
We add the uncertainties in quadrature. 

\begin{figure}[t]
\includegraphics[width=0.45\textwidth]{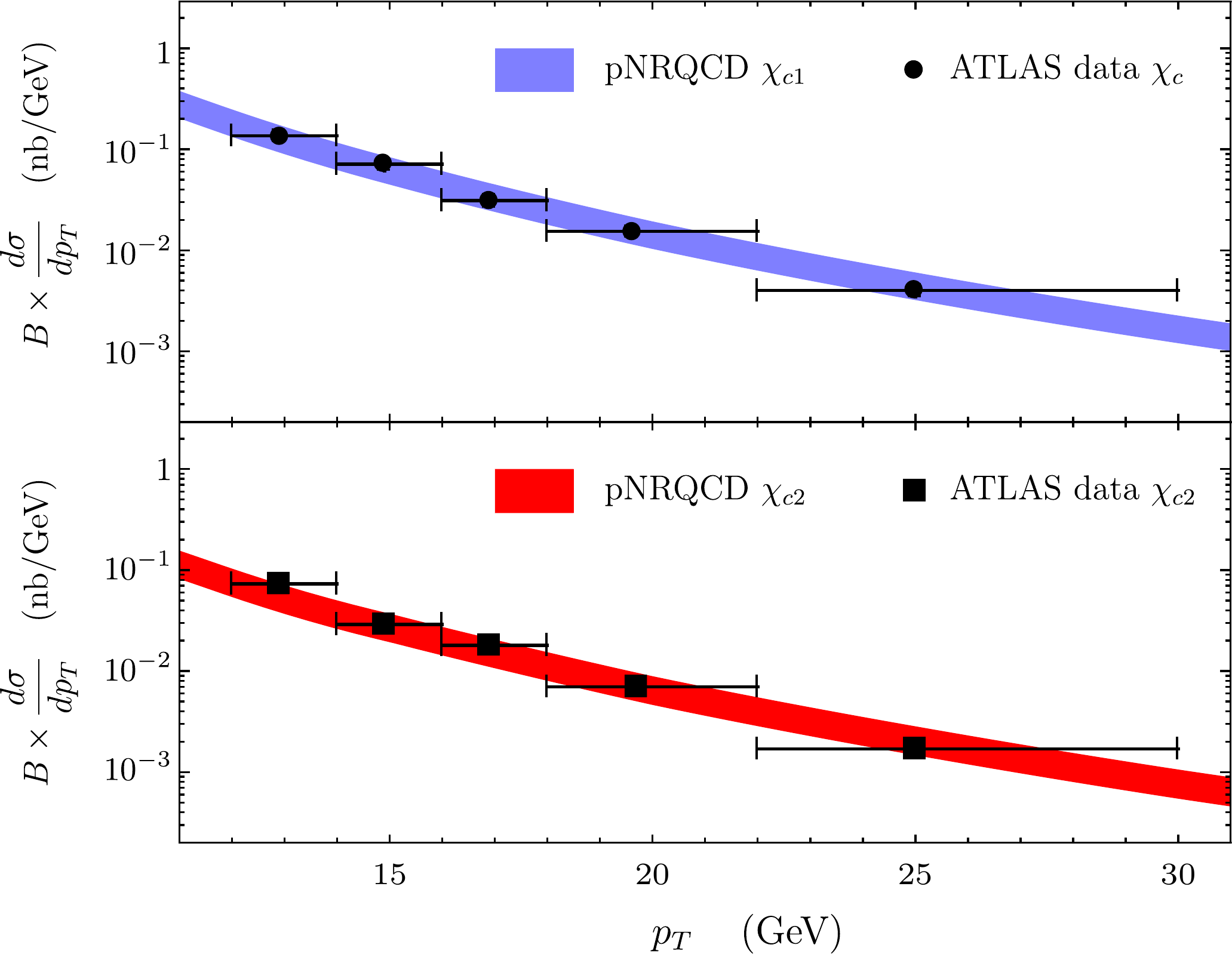}
\caption{\label{fig:chic}
Differential cross sections for $\chi_{c1}$ and $\chi_{c2}$ at the LHC ($\sqrt{s}=7$~TeV, $|y|<0.75$), compared with the ATLAS measurements~\cite{ATLAS:2014ala}.}
\end{figure}

We fit ${\cal E}$ on $\chi_{cJ}$ production data.
In particular, we compute production cross sections of $\chi_{c1}$ and $\chi_{c2}$ by taking the SDCs in Ref.~\cite{Bodwin:2015iua}
that are accurate at NLO in $\alpha_s$ and include resummed leading logarithms of $p_T/m$ for $\sqrt{s}=7$~TeV in the rapidity range $|y|<0.75$. 
Here, $p_T$ is the transverse momentum of the $\chi_{cJ}$.
We present the differential cross sections $B \times d \sigma/d p_T$,
where $B= {\rm Br} (\chi_{cJ} \to J/\psi \gamma) \times {\rm Br} (J/\psi \to \mu^+ \mu^-)$ is taken from Ref.~\cite{Tanabashi:2018oca}.
Our results for the differential cross sections are shown in Fig.~\ref{fig:chic} against ATLAS data~\cite{ATLAS:2014ala}.
The resulting fitted value of ${\cal E}$ in the $\overline{\rm MS}$ scheme is 
\begin{equation}
\label{eq:correlatorproduction}
{\cal E}(\Lambda = 1.5\textrm{~GeV}) = 1.94\pm 0.04.
\end{equation}
The fit quality is good.
We note that the value \eqref{eq:correlatorproduction} is close to ${\cal E}_3 (\Lambda = 1.5\textrm{~GeV}) = 2.73 {}^{+0.94}_{-0.65}$ obtained from $\chi_{cJ}$ decay data~\cite{Brambilla:2020xod}.
This, together with the one-loop running agreement, is suggestive that for the considered $P$-wave CO LDME at leading order in $v$
an approximate crossing relation may hold. 
We note that our results for the $\chi_{cJ}$ LDMEs and the cross sections 
are compatible within uncertainties with the results in Refs.~\cite{Ma:2010vd,
Gong:2012ug, Bodwin:2015iua}.

\begin{figure}[t]
\includegraphics[width=0.45\textwidth]{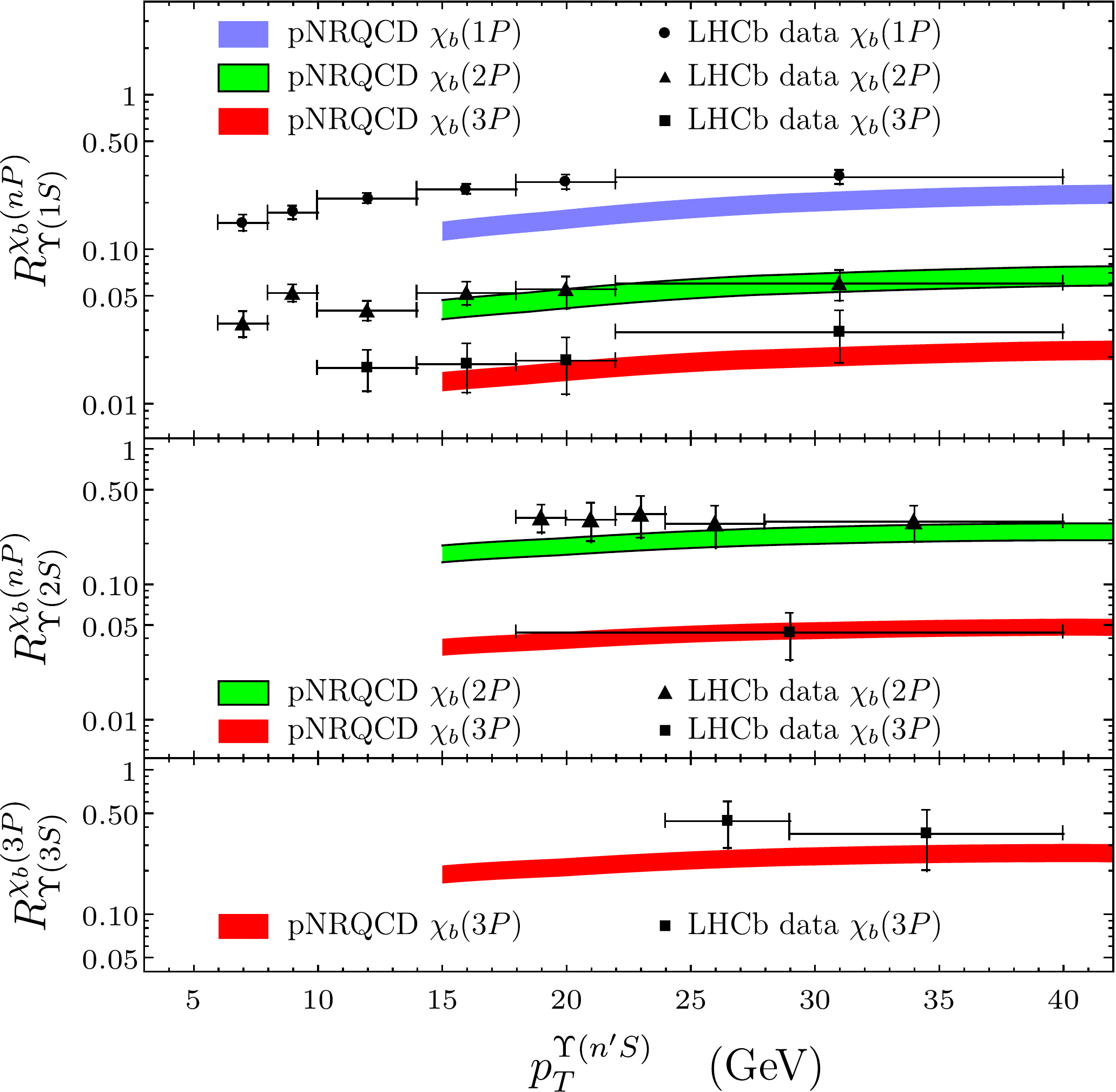}
\caption{\label{fig:RUpschi}
pNRQCD determinations of the feed-down fractions $R^{\chi_b (n P)}_{\Upsilon(n'S)}$ at the LHC ($\sqrt{s}=7$~TeV, $2 < y < 4.5$), compared with the LHCb data~\cite{Aaij:2014caa}.}
\end{figure}

In order to compare with data also in the bottomonium case, we compute the
feed-down fractions $R^{\chi_b (n P)}_{\Upsilon(n'S)}$,
which are defined as the fractions of $\Upsilon(n'S)$ ($n'=1$, 2, and 3) originating from decays of $\chi_{b1} (nP)$ and $\chi_{b2} (nP)$ ($n=$1, 2, and 3). 
We take the correlator ${\cal E}$ extracted from the charmonium production data and run it to the bottom mass at one-loop accuracy.
We compute the SDCs at NLO in $\alpha_s$ for $\sqrt{s}=7$~TeV in the rapidity
range $2 < y < 4.5$ by using the \textsc{fdchqhp} package~\cite{Wan:2014vka}. 
We compute $R^{\chi_b (n P)}_{\Upsilon(n'S)}$ by multiplying the $\chi_{bJ}(nP)$ cross section by the branching ratio ${\rm BR}[\chi_{bJ}(nP) \to \Upsilon(n'S) + \gamma]$,
summing over $J=1$ and $2$, and dividing by the inclusive $\Upsilon(n'S)$ cross section. 
For the branching ratios of the $1P$ and $2P$ states, we take the measured values from Ref.~\cite{Tanabashi:2018oca}, 
while, for the branching ratios of the $3P$ states, we take the theoretical predictions from Ref.~\cite{Han:2014kxa} because there are no data available. 
We present our results as functions of the transverse momentum $p_T^{\Upsilon(n'S)}$ of the $\Upsilon(n'S)$, which we compute 
by multiplying the transverse momentum of the $\chi_{bJ}(nP)$ by $m_{\Upsilon(n'S)}/m_{\chi_{bJ}(nP)}$.
We take the bottomonium masses from Ref.~\cite{Tanabashi:2018oca}.  
For the denominator, we take the $\Upsilon(n'S)$ LDMEs determined in Ref.~\cite{Han:2014kxa}, which give good descriptions of the measured cross sections at the LHC for $p_T > 15$~GeV. 
Our results for $R^{\chi_b (n P)}_{\Upsilon(n'S)}$ are shown in Fig.~\ref{fig:RUpschi} against the LHCb data at $\sqrt{s}=7$~TeV~\cite{Aaij:2014caa}.
Our determinations are in good agreement with the data, with the possible exception of the $\chi_{b}(1P)$ case,
which, however, may not be well described as a strongly coupled quarkonium (see e.g.~\cite{Segovia:2018qzb}).
Our results for the $\chi_{bJ}$ LDMEs and the feed-down fractions 
agree within uncertainties with the results in Ref.~\cite{Han:2014kxa}, while
they disagree with the results in Ref.~\cite{Gong:2013qka}, which were 
obtained from fits to $\Upsilon$ cross section data instead of from 
measurements of $\chi_{bJ}$ production rates.

Our determinations have been made possible by Eq.~\eqref{eq:octet_result}, with ${\cal E}$ fixed on $\chi_{cJ}$ production data,
while previous determinations, including the ones in Ref.~\cite{Han:2014kxa},
relied on using measured bottomonium cross section data to determine the $\chi_{bJ}(nP)$ LDMEs for each $n$. 

The pNRQCD approach for inclusive production of $P$-wave heavy quarkonia that we have developed in this Letter
provides expressions for the color-singlet and color-octet NRQCD LDMEs in terms
of quarkonium wave functions and universal gluonic correlators,
which may be determined from data and/or lattice QCD calculations.
This brings in a reduction in the number of nonperturbative unknowns and a substantial enhancement in the predictive power of the nonrelativistic effective field theory approach
to inclusive heavy quarkonium production processes.
Our results have made possible for the first time to determine the- $\chi_{bJ}(nP)$ cross sections from first principles without fitting color-octet LDMEs on bottomonium production data.
Our results for the $\chi_{cJ}$ cross sections and $\chi_{b}(nP)$ feed-down fractions at the LHC have been summarized in Figs.~\ref{fig:chic} and \ref{fig:RUpschi}. 
They are in good agreement with measurements. 
We note that the formalism developed in this work may be applied to all heavy
quarkonium production processes and, in particular, to the production of the
$J/\psi$ or the $\Upsilon(nS)$ states~\cite{Brambilla:2020long}. 
Similar to our results for the $\chi_{QJ}$ LDMEs in this Letter, 
we expect to find universal, flavor-independent relations between LDMEs for
these states.
It may be also possible to extend the formalism to inclusive production of heavy hybrids and mixed states of hybrids and quarkonia.

\medskip

\begin{acknowledgments}
We are grateful to Geoffrey Bodwin for pointing out an inconsistency in an early version of this work.
The work of N.~B. is supported by the DFG (Deutsche Forschungsgemeinschaft, German Research Foundation) Grant No. BR 4058/2-2.
N.~B., H.~S.~C. and A.~V. acknowledge support from the DFG cluster of excellence ``ORIGINS'' under Germany's Excellence Strategy - EXC-2094 - 390783311.
The work of A. V. is also supported by the DFG and the NSFC (National Science Foundation of China) through funds provided to the Sino-German CRC 110 ``Symmetries and the Emergence of Structure in QCD'' (NSFC Grant No. 11261130311).
\end{acknowledgments}

\bibliography{hadropro-Pwave.bib}

\begin{thebibliography}{40}%
\makeatletter
\providecommand \@ifxundefined [1]{%
 \@ifx{#1\undefined}
}%
\providecommand \@ifnum [1]{%
 \ifnum #1\expandafter \@firstoftwo
 \else \expandafter \@secondoftwo
 \fi
}%
\providecommand \@ifx [1]{%
 \ifx #1\expandafter \@firstoftwo
 \else \expandafter \@secondoftwo
 \fi
}%
\providecommand \natexlab [1]{#1}%
\providecommand \enquote  [1]{``#1''}%
\providecommand \bibnamefont  [1]{#1}%
\providecommand \bibfnamefont [1]{#1}%
\providecommand \citenamefont [1]{#1}%
\providecommand \href@noop [0]{\@secondoftwo}%
\providecommand \href [0]{\begingroup \@sanitize@url \@href}%
\providecommand \@href[1]{\@@startlink{#1}\@@href}%
\providecommand \@@href[1]{\endgroup#1\@@endlink}%
\providecommand \@sanitize@url [0]{\catcode `\\12\catcode `\$12\catcode
  `\&12\catcode `\#12\catcode `\^12\catcode `\_12\catcode `\%12\relax}%
\providecommand \@@startlink[1]{}%
\providecommand \@@endlink[0]{}%
\providecommand \url  [0]{\begingroup\@sanitize@url \@url }%
\providecommand \@url [1]{\endgroup\@href {#1}{\urlprefix }}%
\providecommand \urlprefix  [0]{URL }%
\providecommand \Eprint [0]{\href }%
\providecommand \doibase [0]{https://doi.org/}%
\providecommand \selectlanguage [0]{\@gobble}%
\providecommand \bibinfo  [0]{\@secondoftwo}%
\providecommand \bibfield  [0]{\@secondoftwo}%
\providecommand \translation [1]{[#1]}%
\providecommand \BibitemOpen [0]{}%
\providecommand \bibitemStop [0]{}%
\providecommand \bibitemNoStop [0]{.\EOS\space}%
\providecommand \EOS [0]{\spacefactor3000\relax}%
\providecommand \BibitemShut  [1]{\csname bibitem#1\endcsname}%
\let\auto@bib@innerbib\@empty
\bibitem [{\citenamefont {Brambilla}\ \emph {et~al.}(2004)\citenamefont
  {Brambilla} \emph {et~al.}}]{Brambilla:2004wf}%
  \BibitemOpen
  \bibfield  {author} {\bibinfo {author} {\bibfnamefont {N.}~\bibnamefont
  {Brambilla}} \emph {et~al.} (\bibinfo {collaboration} {Quarkonium Working
  Group}),\ }\bibfield  {title} {\bibinfo {title} {{Heavy quarkonium physics}}\
  }\href {https://doi.org/10.5170/CERN-2005-005} {10.5170/CERN-2005-005}
  (\bibinfo {year} {2004}),\ \Eprint {https://arxiv.org/abs/hep-ph/0412158}
  {arXiv:hep-ph/0412158} \BibitemShut {NoStop}%
\bibitem [{\citenamefont {Brambilla}\ \emph {et~al.}(2011)\citenamefont
  {Brambilla} \emph {et~al.}}]{Brambilla:2010cs}%
  \BibitemOpen
  \bibfield  {author} {\bibinfo {author} {\bibfnamefont {N.}~\bibnamefont
  {Brambilla}} \emph {et~al.},\ }\bibfield  {title} {\bibinfo {title} {{Heavy
  Quarkonium: Progress, Puzzles, and Opportunities}},\ }\href
  {https://doi.org/10.1140/epjc/s10052-010-1534-9} {\bibfield  {journal}
  {\bibinfo  {journal} {Eur. Phys. J. C}\ }\textbf {\bibinfo {volume} {71}},\
  \bibinfo {pages} {1534} (\bibinfo {year} {2011})},\ \Eprint
  {https://arxiv.org/abs/1010.5827} {arXiv:1010.5827 [hep-ph]} \BibitemShut
  {NoStop}%
\bibitem [{\citenamefont {Bodwin}\ \emph {et~al.}(2013)\citenamefont {Bodwin},
  \citenamefont {Braaten}, \citenamefont {Eichten}, \citenamefont {Olsen},
  \citenamefont {Pedlar},\ and\ \citenamefont {Russ}}]{Bodwin:2013nua}%
  \BibitemOpen
  \bibfield  {author} {\bibinfo {author} {\bibfnamefont {G.~T.}\ \bibnamefont
  {Bodwin}}, \bibinfo {author} {\bibfnamefont {E.}~\bibnamefont {Braaten}},
  \bibinfo {author} {\bibfnamefont {E.}~\bibnamefont {Eichten}}, \bibinfo
  {author} {\bibfnamefont {S.~L.}\ \bibnamefont {Olsen}}, \bibinfo {author}
  {\bibfnamefont {T.~K.}\ \bibnamefont {Pedlar}},\ and\ \bibinfo {author}
  {\bibfnamefont {J.}~\bibnamefont {Russ}},\ }\bibfield  {title} {\bibinfo
  {title} {{Quarkonium at the Frontiers of High Energy Physics: A Snowmass
  White Paper}},\ }in\ \href@noop {} {\emph {\bibinfo {booktitle} {{Community
  Summer Study 2013}: {Snowmass on the Mississippi}}}}\ (\bibinfo {year}
  {2013})\ \Eprint {https://arxiv.org/abs/1307.7425} {arXiv:1307.7425 [hep-ph]}
  \BibitemShut {NoStop}%
\bibitem [{\citenamefont {Brambilla}\ \emph {et~al.}(2014)\citenamefont
  {Brambilla} \emph {et~al.}}]{Brambilla:2014jmp}%
  \BibitemOpen
  \bibfield  {author} {\bibinfo {author} {\bibfnamefont {N.}~\bibnamefont
  {Brambilla}} \emph {et~al.},\ }\bibfield  {title} {\bibinfo {title} {{QCD and
  Strongly Coupled Gauge Theories: Challenges and Perspectives}},\ }\href
  {https://doi.org/10.1140/epjc/s10052-014-2981-5} {\bibfield  {journal}
  {\bibinfo  {journal} {Eur. Phys. J. C}\ }\textbf {\bibinfo {volume} {74}},\
  \bibinfo {pages} {2981} (\bibinfo {year} {2014})},\ \Eprint
  {https://arxiv.org/abs/1404.3723} {arXiv:1404.3723 [hep-ph]} \BibitemShut
  {NoStop}%
\bibitem [{\citenamefont {Lansberg}(2019)}]{Lansberg:2019adr}%
  \BibitemOpen
  \bibfield  {author} {\bibinfo {author} {\bibfnamefont {J.-P.}\ \bibnamefont
  {Lansberg}},\ }\bibfield  {title} {\bibinfo {title} {{New Observables in
  Inclusive Production of Quarkonia}},\ }\href@noop {} {\  (\bibinfo {year}
  {2019})},\ \Eprint {https://arxiv.org/abs/1903.09185} {arXiv:1903.09185
  [hep-ph]} \BibitemShut {NoStop}%
\bibitem [{\citenamefont {Matsui}\ and\ \citenamefont
  {Satz}(1986)}]{Matsui:1986dk}%
  \BibitemOpen
  \bibfield  {author} {\bibinfo {author} {\bibfnamefont {T.}~\bibnamefont
  {Matsui}}\ and\ \bibinfo {author} {\bibfnamefont {H.}~\bibnamefont {Satz}},\
  }\bibfield  {title} {\bibinfo {title} {{$J/\psi$ Suppression by Quark-Gluon
  Plasma Formation}},\ }\href {https://doi.org/10.1016/0370-2693(86)91404-8}
  {\bibfield  {journal} {\bibinfo  {journal} {Phys. Lett. B}\ }\textbf
  {\bibinfo {volume} {178}},\ \bibinfo {pages} {416} (\bibinfo {year}
  {1986})}\BibitemShut {NoStop}%
\bibitem [{\citenamefont {Bodwin}\ \emph {et~al.}(1995)\citenamefont {Bodwin},
  \citenamefont {Braaten},\ and\ \citenamefont {Lepage}}]{Bodwin:1994jh}%
  \BibitemOpen
  \bibfield  {author} {\bibinfo {author} {\bibfnamefont {G.~T.}\ \bibnamefont
  {Bodwin}}, \bibinfo {author} {\bibfnamefont {E.}~\bibnamefont {Braaten}},\
  and\ \bibinfo {author} {\bibfnamefont {G.~P.}\ \bibnamefont {Lepage}},\
  }\bibfield  {title} {\bibinfo {title} {{Rigorous QCD analysis of inclusive
  annihilation and production of heavy quarkonium}},\ }\href
  {https://doi.org/10.1103/PhysRevD.55.5853, 10.1103/PhysRevD.51.1125}
  {\bibfield  {journal} {\bibinfo  {journal} {Phys. Rev.}\ }\textbf {\bibinfo
  {volume} {D51}},\ \bibinfo {pages} {1125} (\bibinfo {year} {1995})},\
  \bibinfo {note} {[Erratum: Phys. Rev. D 55, 5853 (1997)]},\ \Eprint
  {https://arxiv.org/abs/hep-ph/9407339} {arXiv:hep-ph/9407339 [hep-ph]}
  \BibitemShut {NoStop}%
\bibitem [{\citenamefont {Nayak}\ \emph
  {et~al.}(2005{\natexlab{a}})\citenamefont {Nayak}, \citenamefont {Qiu},\ and\
  \citenamefont {Sterman}}]{Nayak:2005rw}%
  \BibitemOpen
  \bibfield  {author} {\bibinfo {author} {\bibfnamefont {G.~C.}\ \bibnamefont
  {Nayak}}, \bibinfo {author} {\bibfnamefont {J.-W.}\ \bibnamefont {Qiu}},\
  and\ \bibinfo {author} {\bibfnamefont {G.~F.}\ \bibnamefont {Sterman}},\
  }\bibfield  {title} {\bibinfo {title} {{Fragmentation, factorization and
  infrared poles in heavy quarkonium production}},\ }\href
  {https://doi.org/10.1016/j.physletb.2005.03.031} {\bibfield  {journal}
  {\bibinfo  {journal} {Phys. Lett. B}\ }\textbf {\bibinfo {volume} {613}},\
  \bibinfo {pages} {45} (\bibinfo {year} {2005}{\natexlab{a}})},\ \Eprint
  {https://arxiv.org/abs/hep-ph/0501235} {arXiv:hep-ph/0501235} \BibitemShut
  {NoStop}%
\bibitem [{\citenamefont {Nayak}\ \emph
  {et~al.}(2005{\natexlab{b}})\citenamefont {Nayak}, \citenamefont {Qiu},\ and\
  \citenamefont {Sterman}}]{Nayak:2005rt}%
  \BibitemOpen
  \bibfield  {author} {\bibinfo {author} {\bibfnamefont {G.~C.}\ \bibnamefont
  {Nayak}}, \bibinfo {author} {\bibfnamefont {J.-W.}\ \bibnamefont {Qiu}},\
  and\ \bibinfo {author} {\bibfnamefont {G.~F.}\ \bibnamefont {Sterman}},\
  }\bibfield  {title} {\bibinfo {title} {{Fragmentation, NRQCD and NNLO
  factorization analysis in heavy quarkonium production}},\ }\href
  {https://doi.org/10.1103/PhysRevD.72.114012} {\bibfield  {journal} {\bibinfo
  {journal} {Phys. Rev. D}\ }\textbf {\bibinfo {volume} {72}},\ \bibinfo
  {pages} {114012} (\bibinfo {year} {2005}{\natexlab{b}})},\ \Eprint
  {https://arxiv.org/abs/hep-ph/0509021} {arXiv:hep-ph/0509021} \BibitemShut
  {NoStop}%
\bibitem [{\citenamefont {Nayak}\ \emph {et~al.}(2006)\citenamefont {Nayak},
  \citenamefont {Qiu},\ and\ \citenamefont {Sterman}}]{Nayak:2006fm}%
  \BibitemOpen
  \bibfield  {author} {\bibinfo {author} {\bibfnamefont {G.~C.}\ \bibnamefont
  {Nayak}}, \bibinfo {author} {\bibfnamefont {J.-W.}\ \bibnamefont {Qiu}},\
  and\ \bibinfo {author} {\bibfnamefont {G.~F.}\ \bibnamefont {Sterman}},\
  }\bibfield  {title} {\bibinfo {title} {{NRQCD Factorization and
  Velocity-dependence of NNLO Poles in Heavy Quarkonium Production}},\ }\href
  {https://doi.org/10.1103/PhysRevD.74.074007} {\bibfield  {journal} {\bibinfo
  {journal} {Phys. Rev. D}\ }\textbf {\bibinfo {volume} {74}},\ \bibinfo
  {pages} {074007} (\bibinfo {year} {2006})},\ \Eprint
  {https://arxiv.org/abs/hep-ph/0608066} {arXiv:hep-ph/0608066} \BibitemShut
  {NoStop}%
\bibitem [{\citenamefont {Bodwin}\ \emph {et~al.}(2020)\citenamefont {Bodwin},
  \citenamefont {Chung}, \citenamefont {Ee}, \citenamefont {Kim},\ and\
  \citenamefont {Lee}}]{Bodwin:2019bpf}%
  \BibitemOpen
  \bibfield  {author} {\bibinfo {author} {\bibfnamefont {G.~T.}\ \bibnamefont
  {Bodwin}}, \bibinfo {author} {\bibfnamefont {H.~S.}\ \bibnamefont {Chung}},
  \bibinfo {author} {\bibfnamefont {J.-H.}\ \bibnamefont {Ee}}, \bibinfo
  {author} {\bibfnamefont {U.-R.}\ \bibnamefont {Kim}},\ and\ \bibinfo {author}
  {\bibfnamefont {J.}~\bibnamefont {Lee}},\ }\bibfield  {title} {\bibinfo
  {title} {{Covariant calculation of a two-loop test of nonrelativistic QCD
  factorization}},\ }\href {https://doi.org/10.1103/PhysRevD.101.096011}
  {\bibfield  {journal} {\bibinfo  {journal} {Phys. Rev. D}\ }\textbf {\bibinfo
  {volume} {101}},\ \bibinfo {pages} {096011} (\bibinfo {year} {2020})},\
  \Eprint {https://arxiv.org/abs/1910.05497} {arXiv:1910.05497 [hep-ph]}
  \BibitemShut {NoStop}%
\bibitem [{\citenamefont {Caswell}\ and\ \citenamefont
  {Lepage}(1986)}]{Caswell:1985ui}%
  \BibitemOpen
  \bibfield  {author} {\bibinfo {author} {\bibfnamefont {W.~E.}\ \bibnamefont
  {Caswell}}\ and\ \bibinfo {author} {\bibfnamefont {G.~P.}\ \bibnamefont
  {Lepage}},\ }\bibfield  {title} {\bibinfo {title} {{Effective Lagrangians for
  Bound State Problems in QED, QCD, and Other Field Theories}},\ }\href
  {https://doi.org/10.1016/0370-2693(86)91297-9} {\bibfield  {journal}
  {\bibinfo  {journal} {Phys. Lett.}\ }\textbf {\bibinfo {volume} {167B}},\
  \bibinfo {pages} {437} (\bibinfo {year} {1986})}\BibitemShut {NoStop}%
\bibitem [{\citenamefont {Pineda}\ and\ \citenamefont
  {Soto}(1998)}]{Pineda:1997bj}%
  \BibitemOpen
  \bibfield  {author} {\bibinfo {author} {\bibfnamefont {A.}~\bibnamefont
  {Pineda}}\ and\ \bibinfo {author} {\bibfnamefont {J.}~\bibnamefont {Soto}},\
  }\bibfield  {title} {\bibinfo {title} {{Effective field theory for ultrasoft
  momenta in NRQCD and NRQED}},\ }\href
  {https://doi.org/10.1016/S0920-5632(97)01102-X} {\bibfield  {journal}
  {\bibinfo  {journal} {Nucl. Phys. B Proc. Suppl.}\ }\textbf {\bibinfo
  {volume} {64}},\ \bibinfo {pages} {428} (\bibinfo {year} {1998})},\ \Eprint
  {https://arxiv.org/abs/hep-ph/9707481} {arXiv:hep-ph/9707481} \BibitemShut
  {NoStop}%
\bibitem [{\citenamefont {Brambilla}\ \emph {et~al.}(2000)\citenamefont
  {Brambilla}, \citenamefont {Pineda}, \citenamefont {Soto},\ and\
  \citenamefont {Vairo}}]{Brambilla:1999xf}%
  \BibitemOpen
  \bibfield  {author} {\bibinfo {author} {\bibfnamefont {N.}~\bibnamefont
  {Brambilla}}, \bibinfo {author} {\bibfnamefont {A.}~\bibnamefont {Pineda}},
  \bibinfo {author} {\bibfnamefont {J.}~\bibnamefont {Soto}},\ and\ \bibinfo
  {author} {\bibfnamefont {A.}~\bibnamefont {Vairo}},\ }\bibfield  {title}
  {\bibinfo {title} {{Potential NRQCD: An Effective theory for heavy
  quarkonium}},\ }\href {https://doi.org/10.1016/S0550-3213(99)00693-8}
  {\bibfield  {journal} {\bibinfo  {journal} {Nucl. Phys.}\ }\textbf {\bibinfo
  {volume} {B566}},\ \bibinfo {pages} {275} (\bibinfo {year} {2000})},\ \Eprint
  {https://arxiv.org/abs/hep-ph/9907240} {arXiv:hep-ph/9907240 [hep-ph]}
  \BibitemShut {NoStop}%
\bibitem [{\citenamefont {Brambilla}\ \emph {et~al.}(2005)\citenamefont
  {Brambilla}, \citenamefont {Pineda}, \citenamefont {Soto},\ and\
  \citenamefont {Vairo}}]{Brambilla:2004jw}%
  \BibitemOpen
  \bibfield  {author} {\bibinfo {author} {\bibfnamefont {N.}~\bibnamefont
  {Brambilla}}, \bibinfo {author} {\bibfnamefont {A.}~\bibnamefont {Pineda}},
  \bibinfo {author} {\bibfnamefont {J.}~\bibnamefont {Soto}},\ and\ \bibinfo
  {author} {\bibfnamefont {A.}~\bibnamefont {Vairo}},\ }\bibfield  {title}
  {\bibinfo {title} {{Effective Field Theories for Heavy Quarkonium}},\ }\href
  {https://doi.org/10.1103/RevModPhys.77.1423} {\bibfield  {journal} {\bibinfo
  {journal} {Rev. Mod. Phys.}\ }\textbf {\bibinfo {volume} {77}},\ \bibinfo
  {pages} {1423} (\bibinfo {year} {2005})},\ \Eprint
  {https://arxiv.org/abs/hep-ph/0410047} {arXiv:hep-ph/0410047 [hep-ph]}
  \BibitemShut {NoStop}%
\bibitem [{\citenamefont {Chung}(2018)}]{Chung:2018lyq}%
  \BibitemOpen
  \bibfield  {author} {\bibinfo {author} {\bibfnamefont {H.~S.}\ \bibnamefont
  {Chung}},\ }\bibfield  {title} {\bibinfo {title} {{Review of quarkonium
  production: status and prospects}},\ }\href
  {https://doi.org/10.22323/1.336.0007} {\bibfield  {journal} {\bibinfo
  {journal} {PoS}\ }\textbf {\bibinfo {volume} {Confinement2018}},\ \bibinfo
  {pages} {007} (\bibinfo {year} {2018})},\ \Eprint
  {https://arxiv.org/abs/1811.12098} {arXiv:1811.12098 [hep-ph]} \BibitemShut
  {NoStop}%
\bibitem [{\citenamefont {Brambilla}\ \emph {et~al.}(2002)\citenamefont
  {Brambilla}, \citenamefont {Eiras}, \citenamefont {Pineda}, \citenamefont
  {Soto},\ and\ \citenamefont {Vairo}}]{Brambilla:2001xy}%
  \BibitemOpen
  \bibfield  {author} {\bibinfo {author} {\bibfnamefont {N.}~\bibnamefont
  {Brambilla}}, \bibinfo {author} {\bibfnamefont {D.}~\bibnamefont {Eiras}},
  \bibinfo {author} {\bibfnamefont {A.}~\bibnamefont {Pineda}}, \bibinfo
  {author} {\bibfnamefont {J.}~\bibnamefont {Soto}},\ and\ \bibinfo {author}
  {\bibfnamefont {A.}~\bibnamefont {Vairo}},\ }\bibfield  {title} {\bibinfo
  {title} {{New predictions for inclusive heavy quarkonium $P$ wave decays}},\
  }\href {https://doi.org/10.1103/PhysRevLett.88.012003} {\bibfield  {journal}
  {\bibinfo  {journal} {Phys. Rev. Lett.}\ }\textbf {\bibinfo {volume} {88}},\
  \bibinfo {pages} {012003} (\bibinfo {year} {2002})},\ \Eprint
  {https://arxiv.org/abs/hep-ph/0109130} {arXiv:hep-ph/0109130} \BibitemShut
  {NoStop}%
\bibitem [{\citenamefont {Brambilla}\ \emph {et~al.}(2003)\citenamefont
  {Brambilla}, \citenamefont {Eiras}, \citenamefont {Pineda}, \citenamefont
  {Soto},\ and\ \citenamefont {Vairo}}]{Brambilla:2002nu}%
  \BibitemOpen
  \bibfield  {author} {\bibinfo {author} {\bibfnamefont {N.}~\bibnamefont
  {Brambilla}}, \bibinfo {author} {\bibfnamefont {D.}~\bibnamefont {Eiras}},
  \bibinfo {author} {\bibfnamefont {A.}~\bibnamefont {Pineda}}, \bibinfo
  {author} {\bibfnamefont {J.}~\bibnamefont {Soto}},\ and\ \bibinfo {author}
  {\bibfnamefont {A.}~\bibnamefont {Vairo}},\ }\bibfield  {title} {\bibinfo
  {title} {{Inclusive decays of heavy quarkonium to light particles}},\ }\href
  {https://doi.org/10.1103/PhysRevD.67.034018} {\bibfield  {journal} {\bibinfo
  {journal} {Phys. Rev. D}\ }\textbf {\bibinfo {volume} {67}},\ \bibinfo
  {pages} {034018} (\bibinfo {year} {2003})},\ \Eprint
  {https://arxiv.org/abs/hep-ph/0208019} {arXiv:hep-ph/0208019} \BibitemShut
  {NoStop}%
\bibitem [{\citenamefont {Brambilla}\ \emph {et~al.}(2020)\citenamefont
  {Brambilla}, \citenamefont {Chung}, \citenamefont {M{\"u}ller},\ and\
  \citenamefont {Vairo}}]{Brambilla:2020xod}%
  \BibitemOpen
  \bibfield  {author} {\bibinfo {author} {\bibfnamefont {N.}~\bibnamefont
  {Brambilla}}, \bibinfo {author} {\bibfnamefont {H.~S.}\ \bibnamefont
  {Chung}}, \bibinfo {author} {\bibfnamefont {D.}~\bibnamefont {M{\"u}ller}},\
  and\ \bibinfo {author} {\bibfnamefont {A.}~\bibnamefont {Vairo}},\ }\bibfield
   {title} {\bibinfo {title} {{Decay and electromagnetic production of strongly
  coupled quarkonia in pNRQCD}},\ }\href
  {https://doi.org/10.1007/JHEP04(2020)095} {\bibfield  {journal} {\bibinfo
  {journal} {JHEP}\ }\textbf {\bibinfo {volume} {04}},\ \bibinfo {pages}
  {095}},\ \Eprint {https://arxiv.org/abs/2002.07462} {arXiv:2002.07462
  [hep-ph]} \BibitemShut {NoStop}%
\bibitem [{\citenamefont {Brambilla}\ \emph {et~al.}(2001)\citenamefont
  {Brambilla}, \citenamefont {Pineda}, \citenamefont {Soto},\ and\
  \citenamefont {Vairo}}]{Brambilla:2000gk}%
  \BibitemOpen
  \bibfield  {author} {\bibinfo {author} {\bibfnamefont {N.}~\bibnamefont
  {Brambilla}}, \bibinfo {author} {\bibfnamefont {A.}~\bibnamefont {Pineda}},
  \bibinfo {author} {\bibfnamefont {J.}~\bibnamefont {Soto}},\ and\ \bibinfo
  {author} {\bibfnamefont {A.}~\bibnamefont {Vairo}},\ }\bibfield  {title}
  {\bibinfo {title} {{The QCD potential at $O(1/m)$}},\ }\href
  {https://doi.org/10.1103/PhysRevD.63.014023} {\bibfield  {journal} {\bibinfo
  {journal} {Phys. Rev. D}\ }\textbf {\bibinfo {volume} {63}},\ \bibinfo
  {pages} {014023} (\bibinfo {year} {2001})},\ \Eprint
  {https://arxiv.org/abs/hep-ph/0002250} {arXiv:hep-ph/0002250} \BibitemShut
  {NoStop}%
\bibitem [{\citenamefont {Pineda}\ and\ \citenamefont
  {Vairo}(2001)}]{Pineda:2000sz}%
  \BibitemOpen
  \bibfield  {author} {\bibinfo {author} {\bibfnamefont {A.}~\bibnamefont
  {Pineda}}\ and\ \bibinfo {author} {\bibfnamefont {A.}~\bibnamefont {Vairo}},\
  }\bibfield  {title} {\bibinfo {title} {{The QCD potential at $O(1/m^2)$:
  Complete spin dependent and spin independent result}},\ }\href
  {https://doi.org/10.1103/PhysRevD.64.039902} {\bibfield  {journal} {\bibinfo
  {journal} {Phys. Rev. D}\ }\textbf {\bibinfo {volume} {63}},\ \bibinfo
  {pages} {054007} (\bibinfo {year} {2001})},\ \bibinfo {note} {[Erratum: Phys.
  Rev. D 64, 039902 (2001)]},\ \Eprint {https://arxiv.org/abs/hep-ph/0009145}
  {arXiv:hep-ph/0009145} \BibitemShut {NoStop}%
\bibitem [{Note1()}]{Note1}%
  \BibitemOpen
  \bibinfo {note} {In the special case $n=0$, it holds that ${\rm Tr}\{| 0; \bm
  {x}_1, \bm {x}_2 \rangle ^{(0)}\} = \sqrt {N_c} |\Omega \rangle $ for
  $\protect \bm {x}_1 -\protect \bm {x}_2 \to \protect \bm {0}$~\cite
  {Brambilla:2002nu,Brambilla:2020xod}.}\BibitemShut {Stop}%
\bibitem [{\citenamefont {Makeenko}\ and\ \citenamefont
  {Migdal}(1979)}]{Makeenko:1979pb}%
  \BibitemOpen
  \bibfield  {author} {\bibinfo {author} {\bibfnamefont {Y.}~\bibnamefont
  {Makeenko}}\ and\ \bibinfo {author} {\bibfnamefont {A.~A.}\ \bibnamefont
  {Migdal}},\ }\bibfield  {title} {\bibinfo {title} {{Exact Equation for the
  Loop Average in Multicolor QCD}},\ }\href
  {https://doi.org/10.1016/0370-2693(79)90131-X} {\bibfield  {journal}
  {\bibinfo  {journal} {Phys. Lett. B}\ }\textbf {\bibinfo {volume} {88}},\
  \bibinfo {pages} {135} (\bibinfo {year} {1979})},\ \bibinfo {note} {[Erratum:
  Phys. Lett. B 89, 437 (1980)]}\BibitemShut {NoStop}%
\bibitem [{\citenamefont {Witten}(1980)}]{Witten:1979pi}%
  \BibitemOpen
  \bibfield  {author} {\bibinfo {author} {\bibfnamefont {E.}~\bibnamefont
  {Witten}},\ }\bibfield  {title} {\bibinfo {title} {{The $1/N$ expansion in
  atomic and particle physics}},\ }\href
  {https://doi.org/10.1007/978-1-4684-7571-5\_21} {\bibfield  {journal}
  {\bibinfo  {journal} {NATO Sci. Ser. B}\ }\textbf {\bibinfo {volume} {59}},\
  \bibinfo {pages} {403} (\bibinfo {year} {1980})}\BibitemShut {NoStop}%
\bibitem [{\citenamefont {Ablikim}\ \emph {et~al.}(2012)\citenamefont {Ablikim}
  \emph {et~al.}}]{Ablikim:2012xi}%
  \BibitemOpen
  \bibfield  {author} {\bibinfo {author} {\bibfnamefont {M.}~\bibnamefont
  {Ablikim}} \emph {et~al.} (\bibinfo {collaboration} {BESIII}),\ }\bibfield
  {title} {\bibinfo {title} {{Two-photon widths of the $\chi_{c0, 2}$ states
  and helicity analysis for $\chi_{c2}\to\gamma\gamma$}},\ }\href
  {https://doi.org/10.1103/PhysRevD.85.112008} {\bibfield  {journal} {\bibinfo
  {journal} {Phys. Rev. D}\ }\textbf {\bibinfo {volume} {85}},\ \bibinfo
  {pages} {112008} (\bibinfo {year} {2012})},\ \Eprint
  {https://arxiv.org/abs/1205.4284} {arXiv:1205.4284 [hep-ex]} \BibitemShut
  {NoStop}%
\bibitem [{\citenamefont {Buchm{\"u}ller}\ and\ \citenamefont
  {Tye}(1981)}]{Buchmuller:1980su}%
  \BibitemOpen
  \bibfield  {author} {\bibinfo {author} {\bibfnamefont {W.}~\bibnamefont
  {Buchm{\"u}ller}}\ and\ \bibinfo {author} {\bibfnamefont {S.}~\bibnamefont
  {Tye}},\ }\bibfield  {title} {\bibinfo {title} {{Quarkonia and Quantum
  Chromodynamics}},\ }\href {https://doi.org/10.1103/PhysRevD.24.132}
  {\bibfield  {journal} {\bibinfo  {journal} {Phys. Rev. D}\ }\textbf {\bibinfo
  {volume} {24}},\ \bibinfo {pages} {132} (\bibinfo {year} {1981})}\BibitemShut
  {NoStop}%
\bibitem [{\citenamefont {Eichten}\ and\ \citenamefont
  {Quigg}(1995)}]{Eichten:1995ch}%
  \BibitemOpen
  \bibfield  {author} {\bibinfo {author} {\bibfnamefont {E.~J.}\ \bibnamefont
  {Eichten}}\ and\ \bibinfo {author} {\bibfnamefont {C.}~\bibnamefont
  {Quigg}},\ }\bibfield  {title} {\bibinfo {title} {{Quarkonium wave functions
  at the origin}},\ }\href {https://doi.org/10.1103/PhysRevD.52.1726}
  {\bibfield  {journal} {\bibinfo  {journal} {Phys. Rev. D}\ }\textbf {\bibinfo
  {volume} {52}},\ \bibinfo {pages} {1726} (\bibinfo {year} {1995})},\ \Eprint
  {https://arxiv.org/abs/hep-ph/9503356} {arXiv:hep-ph/9503356} \BibitemShut
  {NoStop}%
\bibitem [{\citenamefont {Chung}\ \emph {et~al.}(2011)\citenamefont {Chung},
  \citenamefont {Lee},\ and\ \citenamefont {Yu}}]{Chung:2010vz}%
  \BibitemOpen
  \bibfield  {author} {\bibinfo {author} {\bibfnamefont {H.~S.}\ \bibnamefont
  {Chung}}, \bibinfo {author} {\bibfnamefont {J.}~\bibnamefont {Lee}},\ and\
  \bibinfo {author} {\bibfnamefont {C.}~\bibnamefont {Yu}},\ }\bibfield
  {title} {\bibinfo {title} {{NRQCD matrix elements for S-wave bottomonia and
  $\Gamma[\eta_b(nS) \to \gamma \gamma]$ with relativistic corrections}},\
  }\href {https://doi.org/10.1016/j.physletb.2011.01.033} {\bibfield  {journal}
  {\bibinfo  {journal} {Phys. Lett. B}\ }\textbf {\bibinfo {volume} {697}},\
  \bibinfo {pages} {48} (\bibinfo {year} {2011})},\ \Eprint
  {https://arxiv.org/abs/1011.1554} {arXiv:1011.1554 [hep-ph]} \BibitemShut
  {NoStop}%
\bibitem [{\citenamefont {Eichten}\ and\ \citenamefont
  {Quigg}(2019)}]{Eichten:2019hbb}%
  \BibitemOpen
  \bibfield  {author} {\bibinfo {author} {\bibfnamefont {E.~J.}\ \bibnamefont
  {Eichten}}\ and\ \bibinfo {author} {\bibfnamefont {C.}~\bibnamefont
  {Quigg}},\ }\bibfield  {title} {\bibinfo {title} {{Quarkonium wave functions
  at the origin: an update}},\ }\href@noop {} {\  (\bibinfo {year} {2019})},\
  \Eprint {https://arxiv.org/abs/1904.11542} {arXiv:1904.11542 [hep-ph]}
  \BibitemShut {NoStop}%
\bibitem [{\citenamefont {Aad}\ \emph {et~al.}(2014)\citenamefont {Aad} \emph
  {et~al.}}]{ATLAS:2014ala}%
  \BibitemOpen
  \bibfield  {author} {\bibinfo {author} {\bibfnamefont {G.}~\bibnamefont
  {Aad}} \emph {et~al.} (\bibinfo {collaboration} {ATLAS}),\ }\bibfield
  {title} {\bibinfo {title} {{Measurement of $\chi_{c1}$ and $\chi_{c2}$
  production with $\sqrt{s}$ = 7 TeV $pp$ collisions at ATLAS}},\ }\href
  {https://doi.org/10.1007/JHEP07(2014)154} {\bibfield  {journal} {\bibinfo
  {journal} {JHEP}\ }\textbf {\bibinfo {volume} {07}},\ \bibinfo {pages}
  {154}},\ \Eprint {https://arxiv.org/abs/1404.7035} {arXiv:1404.7035 [hep-ex]}
  \BibitemShut {NoStop}%
\bibitem [{\citenamefont {Bodwin}\ \emph {et~al.}(2016)\citenamefont {Bodwin},
  \citenamefont {Chao}, \citenamefont {Chung}, \citenamefont {Kim},
  \citenamefont {Lee},\ and\ \citenamefont {Ma}}]{Bodwin:2015iua}%
  \BibitemOpen
  \bibfield  {author} {\bibinfo {author} {\bibfnamefont {G.~T.}\ \bibnamefont
  {Bodwin}}, \bibinfo {author} {\bibfnamefont {K.-T.}\ \bibnamefont {Chao}},
  \bibinfo {author} {\bibfnamefont {H.~S.}\ \bibnamefont {Chung}}, \bibinfo
  {author} {\bibfnamefont {U.-R.}\ \bibnamefont {Kim}}, \bibinfo {author}
  {\bibfnamefont {J.}~\bibnamefont {Lee}},\ and\ \bibinfo {author}
  {\bibfnamefont {Y.-Q.}\ \bibnamefont {Ma}},\ }\bibfield  {title} {\bibinfo
  {title} {{Fragmentation contributions to hadroproduction of prompt $J/\psi$,
  $\chi_{cJ}$, and $\psi(2S)$ states}},\ }\href
  {https://doi.org/10.1103/PhysRevD.93.034041} {\bibfield  {journal} {\bibinfo
  {journal} {Phys. Rev. D}\ }\textbf {\bibinfo {volume} {93}},\ \bibinfo
  {pages} {034041} (\bibinfo {year} {2016})},\ \Eprint
  {https://arxiv.org/abs/1509.07904} {arXiv:1509.07904 [hep-ph]} \BibitemShut
  {NoStop}%
\bibitem [{\citenamefont {Tanabashi}\ \emph {et~al.}(2018)\citenamefont
  {Tanabashi} \emph {et~al.}}]{Tanabashi:2018oca}%
  \BibitemOpen
  \bibfield  {author} {\bibinfo {author} {\bibfnamefont {M.}~\bibnamefont
  {Tanabashi}} \emph {et~al.} (\bibinfo {collaboration} {Particle Data
  Group}),\ }\bibfield  {title} {\bibinfo {title} {{Review of Particle
  Physics}},\ }\href {https://doi.org/10.1103/PhysRevD.98.030001} {\bibfield
  {journal} {\bibinfo  {journal} {Phys. Rev. D}\ }\textbf {\bibinfo {volume}
  {98}},\ \bibinfo {pages} {030001} (\bibinfo {year} {2018})}\BibitemShut
  {NoStop}%
\bibitem [{\citenamefont {Ma}\ \emph {et~al.}(2011)\citenamefont {Ma},
  \citenamefont {Wang},\ and\ \citenamefont {Chao}}]{Ma:2010vd}%
  \BibitemOpen
  \bibfield  {author} {\bibinfo {author} {\bibfnamefont {Y.-Q.}\ \bibnamefont
  {Ma}}, \bibinfo {author} {\bibfnamefont {K.}~\bibnamefont {Wang}},\ and\
  \bibinfo {author} {\bibfnamefont {K.-T.}\ \bibnamefont {Chao}},\ }\bibfield
  {title} {\bibinfo {title} {{QCD radiative corrections to $\chi_{cJ}$
  production at hadron colliders}},\ }\href
  {https://doi.org/10.1103/PhysRevD.83.111503} {\bibfield  {journal} {\bibinfo
  {journal} {Phys. Rev. D}\ }\textbf {\bibinfo {volume} {83}},\ \bibinfo
  {pages} {111503} (\bibinfo {year} {2011})},\ \Eprint
  {https://arxiv.org/abs/1002.3987} {arXiv:1002.3987 [hep-ph]} \BibitemShut
  {NoStop}%
\bibitem [{\citenamefont {Gong}\ \emph {et~al.}(2013)\citenamefont {Gong},
  \citenamefont {Wan}, \citenamefont {Wang},\ and\ \citenamefont
  {Zhang}}]{Gong:2012ug}%
  \BibitemOpen
  \bibfield  {author} {\bibinfo {author} {\bibfnamefont {B.}~\bibnamefont
  {Gong}}, \bibinfo {author} {\bibfnamefont {L.-P.}\ \bibnamefont {Wan}},
  \bibinfo {author} {\bibfnamefont {J.-X.}\ \bibnamefont {Wang}},\ and\
  \bibinfo {author} {\bibfnamefont {H.-F.}\ \bibnamefont {Zhang}},\ }\bibfield
  {title} {\bibinfo {title} {{Polarization for Prompt J/\ensuremath{\psi} and
  \ensuremath{\psi}(2s) Production at the Tevatron and LHC}},\ }\href
  {https://doi.org/10.1103/PhysRevLett.110.042002} {\bibfield  {journal}
  {\bibinfo  {journal} {Phys. Rev. Lett.}\ }\textbf {\bibinfo {volume} {110}},\
  \bibinfo {pages} {042002} (\bibinfo {year} {2013})},\ \Eprint
  {https://arxiv.org/abs/1205.6682} {arXiv:1205.6682 [hep-ph]} \BibitemShut
  {NoStop}%
\bibitem [{\citenamefont {Aaij}\ \emph {et~al.}(2014)\citenamefont {Aaij} \emph
  {et~al.}}]{Aaij:2014caa}%
  \BibitemOpen
  \bibfield  {author} {\bibinfo {author} {\bibfnamefont {R.}~\bibnamefont
  {Aaij}} \emph {et~al.} (\bibinfo {collaboration} {LHCb}),\ }\bibfield
  {title} {\bibinfo {title} {{Study of $\chi_b$ meson production in $pp$
  collisions at $\sqrt{s}=7$ and $8{\mathrm {\,TeV}} $ and observation of the
  decay $\chi _b(3P) \to \Upsilon(3S) \gamma$}},\ }\href
  {https://doi.org/10.1140/epjc/s10052-014-3092-z} {\bibfield  {journal}
  {\bibinfo  {journal} {Eur. Phys. J. C}\ }\textbf {\bibinfo {volume} {74}},\
  \bibinfo {pages} {3092} (\bibinfo {year} {2014})},\ \Eprint
  {https://arxiv.org/abs/1407.7734} {arXiv:1407.7734 [hep-ex]} \BibitemShut
  {NoStop}%
\bibitem [{\citenamefont {Wan}\ and\ \citenamefont {Wang}(2014)}]{Wan:2014vka}%
  \BibitemOpen
  \bibfield  {author} {\bibinfo {author} {\bibfnamefont {L.-P.}\ \bibnamefont
  {Wan}}\ and\ \bibinfo {author} {\bibfnamefont {J.-X.}\ \bibnamefont {Wang}},\
  }\bibfield  {title} {\bibinfo {title} {{FDCHQHP: A Fortran package for heavy
  quarkonium hadroproduction}},\ }\href
  {https://doi.org/10.1016/j.cpc.2014.06.022} {\bibfield  {journal} {\bibinfo
  {journal} {Comput. Phys. Commun.}\ }\textbf {\bibinfo {volume} {185}},\
  \bibinfo {pages} {2939} (\bibinfo {year} {2014})},\ \Eprint
  {https://arxiv.org/abs/1405.2143} {arXiv:1405.2143 [hep-ph]} \BibitemShut
  {NoStop}%
\bibitem [{\citenamefont {Han}\ \emph {et~al.}(2016)\citenamefont {Han},
  \citenamefont {Ma}, \citenamefont {Meng}, \citenamefont {Shao}, \citenamefont
  {Zhang},\ and\ \citenamefont {Chao}}]{Han:2014kxa}%
  \BibitemOpen
  \bibfield  {author} {\bibinfo {author} {\bibfnamefont {H.}~\bibnamefont
  {Han}}, \bibinfo {author} {\bibfnamefont {Y.-Q.}\ \bibnamefont {Ma}},
  \bibinfo {author} {\bibfnamefont {C.}~\bibnamefont {Meng}}, \bibinfo {author}
  {\bibfnamefont {H.-S.}\ \bibnamefont {Shao}}, \bibinfo {author}
  {\bibfnamefont {Y.-J.}\ \bibnamefont {Zhang}},\ and\ \bibinfo {author}
  {\bibfnamefont {K.-T.}\ \bibnamefont {Chao}},\ }\bibfield  {title} {\bibinfo
  {title} {{$\Upsilon(nS)$ and $\chi_b(nP)$ production at hadron colliders in
  nonrelativistic QCD}},\ }\href {https://doi.org/10.1103/PhysRevD.94.014028}
  {\bibfield  {journal} {\bibinfo  {journal} {Phys. Rev. D}\ }\textbf {\bibinfo
  {volume} {94}},\ \bibinfo {pages} {014028} (\bibinfo {year} {2016})},\
  \Eprint {https://arxiv.org/abs/1410.8537} {arXiv:1410.8537 [hep-ph]}
  \BibitemShut {NoStop}%
\bibitem [{\citenamefont {Segovia}\ \emph {et~al.}(2019)\citenamefont
  {Segovia}, \citenamefont {Steinbei{\ss}er},\ and\ \citenamefont
  {Vairo}}]{Segovia:2018qzb}%
  \BibitemOpen
  \bibfield  {author} {\bibinfo {author} {\bibfnamefont {J.}~\bibnamefont
  {Segovia}}, \bibinfo {author} {\bibfnamefont {S.}~\bibnamefont
  {Steinbei{\ss}er}},\ and\ \bibinfo {author} {\bibfnamefont {A.}~\bibnamefont
  {Vairo}},\ }\bibfield  {title} {\bibinfo {title} {{Electric dipole
  transitions of $1P$ bottomonia}},\ }\href
  {https://doi.org/10.1103/PhysRevD.99.074011} {\bibfield  {journal} {\bibinfo
  {journal} {Phys. Rev. D}\ }\textbf {\bibinfo {volume} {99}},\ \bibinfo
  {pages} {074011} (\bibinfo {year} {2019})},\ \Eprint
  {https://arxiv.org/abs/1811.07590} {arXiv:1811.07590 [hep-ph]} \BibitemShut
  {NoStop}%
\bibitem [{\citenamefont {Gong}\ \emph {et~al.}(2014)\citenamefont {Gong},
  \citenamefont {Wan}, \citenamefont {Wang},\ and\ \citenamefont
  {Zhang}}]{Gong:2013qka}%
  \BibitemOpen
  \bibfield  {author} {\bibinfo {author} {\bibfnamefont {B.}~\bibnamefont
  {Gong}}, \bibinfo {author} {\bibfnamefont {L.-P.}\ \bibnamefont {Wan}},
  \bibinfo {author} {\bibfnamefont {J.-X.}\ \bibnamefont {Wang}},\ and\
  \bibinfo {author} {\bibfnamefont {H.-F.}\ \bibnamefont {Zhang}},\ }\bibfield
  {title} {\bibinfo {title} {{Complete next-to-leading-order study on the yield
  and polarization of $\Upsilon(1S,2S,3S)$ at the Tevatron and LHC}},\ }\href
  {https://doi.org/10.1103/PhysRevLett.112.032001} {\bibfield  {journal}
  {\bibinfo  {journal} {Phys. Rev. Lett.}\ }\textbf {\bibinfo {volume} {112}},\
  \bibinfo {pages} {032001} (\bibinfo {year} {2014})},\ \Eprint
  {https://arxiv.org/abs/1305.0748} {arXiv:1305.0748 [hep-ph]} \BibitemShut
  {NoStop}%
\bibitem [{\citenamefont {Brambilla}\ \emph {et~al.}()\citenamefont
  {Brambilla}, \citenamefont {Chung},\ and\ \citenamefont
  {Vairo}}]{Brambilla:2020long}%
  \BibitemOpen
  \bibfield  {author} {\bibinfo {author} {\bibfnamefont {N.}~\bibnamefont
  {Brambilla}}, \bibinfo {author} {\bibfnamefont {H.~S.}\ \bibnamefont
  {Chung}},\ and\ \bibinfo {author} {\bibfnamefont {A.}~\bibnamefont {Vairo}},\
  }\href@noop {} {}\bibinfo {howpublished} {in preparation, TUM-EFT
  139/20}\BibitemShut {NoStop}%
\end{thebibliography}%

\end{document}